\documentclass[sn-basic]{sn-jnl}% Basic Springer Nature Reference Style/Chemistry Reference Style
\usepackage{graphicx}%
\usepackage{multirow}%
\usepackage{amsmath,amssymb,amsfonts}%
\usepackage{amsthm}%
\usepackage{mathrsfs}%
\usepackage[title]{appendix}%
\usepackage{xcolor}%
\usepackage{textcomp}%
\usepackage{manyfoot}%
\usepackage{booktabs}%
\usepackage{algorithm}%
\usepackage{algorithmicx}%
\usepackage{algpseudocode}%
\usepackage{listings}%
\usepackage{natbib}%
\usepackage{subfigure}
\usepackage{subcaption}
\usepackage{float}
\usepackage[utf8]{inputenc}
\usepackage{ulem}
\theoremstyle{thmstyleone}%
\theoremstyle{thmstyletwo}%

\theoremstyle{thmstylethree}%

\begin{document}
\title[Article Title]{Realization of a clock-based global height system: A simulation study for Europe and Brazil}


\author*[1]{\fnm{Asha} \sur{Vincent}}\email{vincent@ife.uni-hannover.de}

\author[1]{\fnm{J\"urgen} \sur{M\"uller}}\email{mueller@ife.uni-hannover.de}

\author[2]{\fnm{Christian} \sur{Lisdat}}\email{Christian.Lisdat@ptb.de}

\author[3]{\fnm{Dennis} \sur{Philipp}}\email{dennis.philipp@zarm.uni-bremen.de}

\affil*[1]{\orgdiv{Institute for Geodesy}, \orgname{Leibniz University Hannover}, \orgaddress{\street{Schneiderberg 50}, \city{Hannover}, \postcode{30165}, \state{Lower Saxony}, \country{Germany}}}

\affil[2]{\orgname{Physikalisch-Technische Bundesanstalt}, \orgaddress{\street{Bundesallee 100}, \city{Braunschweig}, \postcode{38116}, \state{Lower Saxony}, \country{Germany}}}

\affil[3]{\orgdiv{ZARM}, \orgname{University of Bremen}, \orgaddress{\street{Am Fallturm 2}, \city{Bremen}, \postcode{28359}, \state{Bremen}, \country{Germany}}}

\abstract{Chronometric levelling is a novel technique for the realisation of the International Height Reference System (IHRS). A detailed study of this technique is carried out through end-to-end simulations, aiming to unify regional/local height systems (LHS) in Europe and Brazil. Focusing on a unification accuracy of 1~cm, realistic scenarios considering an extended parametric model for the vertical datum parameters and detailed clock observation uncertainties were analysed. The vertical datum parameters represent datum offsets, local vertical datum alignment discrepancies in latitude and longitude, accumulated tilts depending on the distance from the reference tide gauge and levelling point elevation-dependent offsets. Clocks achieving a fractional uncertainty of $10^{-18}$ and $10^{-17}$ were assumed in the simulations, considering correlations of clock intrinsic uncertainties, external effects on clock observations such as tidal effects, propagation delay in terms of link uncertainties and outliers. We determine the optimal distribution of clocks across the LHS for the accurate estimation of the vertical datum parameters. We conclude that the estimation of these parameters is related to the spatial distribution of the clocks and recommend an optimal setup placing the clocks at the most distant levelling points, reference tide gauges and elevated points. Further, we investigate the accuracy implications of using an ensemble of clocks based on a master clock and some local clocks per LHS, with the LHS connected only through the master clocks. This results in a configuration with fewer links and fewer clock observations. Our results suggest that a vertical datum unification in Europe and Brazil can be achieved with an accuracy of 1–2~cm. The connection of this unified European/Brazilian height system to the global geoid defined by the conventional $W_0$ value of the IHRS provides geoid-related heights with an uncertainty of about 3~cm.} %With the current portable high-performance clocks of $10^{-18}$ uncertainties clock campaigns can be conducted that make the chronometric levelling less futuristic.}}




\maketitle

\section{Introduction}\label{sec1}

The necessity of a global height system is paramount for various geodetic applications, such as infrastructure development across borders, disaster management, climate studies, etc., as it provides a globally consistent framework \citep{sanchez2021}. Currently, various regional height systems refer to local vertical datums based on the mean sea level determined at an arbitrarily selected reference tide gauge. The heights are then propagated by means of classical levelling techniques or by combining satellite-based positioning with a geoid model \citep{torge2023, delva2019}. Local vertical datums may not coincide with the global geoid, an equipotential surface of the Earth's gravity field that best fits the global mean sea level at rest, due to several factors such as sea surface topography, ocean circulation, local currents and tides \citep{gerlach2013}. Heights and derived quantities, such as gravity anomalies, defined with respect to these local references, are inappropriate for international applications as they cannot be compared unless the biases between the local height systems (LHS) are known. A common vertical reference defined in relation to the global geoid offers global consistency and reliability for the simultaneous determination and monitoring of the geodetic parameters. Hence, an International Height Reference System (IHRS) is in demand, and it can be achieved by unifying the existing LHS worldwide \citep{ihde2017, sanchez2021}. 

Current unification techniques for the IHRS involve integrating spaceborne and terrestrial gravity data, along with precise geodetic positioning using GNSS (Global Navigation Satellite Systems) and advanced satellite gravimetry (e.g., GRACE (Gravity Recovery and Climate Experiment) and GOCE (Gravity Field and Steady-State Ocean Circulation Explorer)) \citep{gerlach2013, sideris2014geodetic, rulke2012unification}. The target accuracy for vertical station positions is \( \pm 3 \, \text{mm} \), and \( \pm 0.03 \, \text{m}^2/\text{s}^2 \) for geopotential differences at the global level, but since these goals are currently unrealistic, efforts focus on static potential values with an achievable accuracy of about 1 cm \citep{sanchez2021}. The use of highly precise atomic clocks, especially optical clocks, offers a promising approach for the realisation of the IHRS by enabling direct, point-wise measurements of gravity potential differences between the IHRS sites \citep{wu2018, wu2020}. This method provides a direct link to geopotential numbers, overcoming the limitations of traditional height determination techniques. By comparing clock frequencies at different sites, gravity potential and height differences can be determined with enhanced accuracy, detecting small variations in near real-time, something that current satellite missions cannot achieve \citep{vincent2023, vincent2024}. An IHRS realization can be established with a proper network of atomic clocks, either by unifying existing LHS by means of clock frequency comparisons or by materialising the reference stations with atomic clocks, i.e., a chronometric reference frame. This study focuses on the unification of LHS using simulated clock observations.

\section{IHRS Realization through Chronometric Levelling}\label{sec2}

According to the IAG convention for the IHRS, the vertical coordinate of a point $P$ is the geopotential number $C_P = W_0 - W_P$, which represents the difference between the gravity potential at the geoid $W_0$ and at the point of interest $W_P$ \citep{drewes2016, sanchez2023}. Here, the positive sign convention is used, in which the geopotential $W$ is always positive and tends to zero at infinite distance from the mass concentration.
Hence, $C_P$ provides the actual potential difference at the point $P$, i.e., it  includes geoid height variation and Earth's surface deformation at the measurement epoch \citep{jekeli2000, vincent2023}. 
A clock comparison determines the difference in elapsed proper times or the relative frequency difference, the redshift, between two clocks. 
The redshift of two standard clocks (i.e., a clock that measures its own elapsed proper time) is affected by the individual positions in the gravity field and the relative state of motion. In the case of two Earth-bound (stationary) clocks, the second part is related to the centrifugal potential.
In the following, we always assume that all involved clocks are indeed standard clocks, as all experiments to date agree that atomic clocks are standard clocks. In geodetic applications, a first-order post-Newtonian (1PN) approximation is commonly adopted, in which relativistic contributions up to order $1/c^2$, with $c$ being the speed of light, are retained. The redshift $z$ or the relative fractional frequency difference between clocks at sites 1 and 2 corresponds to the difference in height $\Delta H$ between the two sites, expressed as,
%Equation
\begin{equation}
z = \frac{\Delta f}{f} \approx \frac{\Delta W}{c^2} \approx \frac{g \Delta H}{c^2} \, ,
\label{eq:z}
\end{equation}
where $f$ represents the frequency, $\Delta W = W_1 - W_2 = C_2 - C_1$ is the corresponding gravity potential difference \citep{bjer1985, muller2018, denker2018, pavlis2003relativistic, pavlis2017re} and $g$ is the gravity acceleration.
In Eq. (\ref{eq:z}), $W$ denotes the gravity potential at the clock's position on the Earth’s surface, defined as the sum of the Newtonian gravitational potential and the centrifugal potential associated with Earth’s rotation. 
Note that a generalized equation is available that holds at all orders in full general relativity, see e.g., \cite{gervsl2015relativistic, wolf1995relativistic, philipp2020}.
For two stationary clocks being virtually on the geoid and at a surface point $P$, respectively, the fractional frequency difference would provide a direct measurement of the geopotential number $C_P$,
\begin{align}
    \label{eq:geoPot2clkObs}
    z = \frac{\Delta f}{f} \approx \dfrac{C_P}{c^2} \approx \dfrac{g H_P}{c^2} \, ,
\end{align}
where $H_P$ is the height of point $P$ above the geoid.
High-performance clocks of fractional uncertainty $10^{-18}$ as already realized in labs, e.g., at PTB (Physikalisch-Technische Bundesanstalt), NIST (National Institute of Standards and Technology), JILA (Joint Institute for Laboratory Astrophysics), or at the National University of Singapore \citep{sanner2019, collaboration2021, arn24}, are sensitive to a potential variation of $0.1\,$m$^2$/s$^2$ which corresponds to a height difference of 1 cm close to the Earth's surface \citep{mcgrew2018,bothwell2022}.

\section{Methodology}\label{sec:method}
We have adopted the approach introduced in \cite{wu2018} and conducted an extended and dedicated set of simulations for a quantitative assessment of the vertical datum unification accuracy under realistic assumptions. For the following, the framework of the first-order post-Newtonian theory for clock comparison is employed. For our simulation studies, an a priori height system, which is considered to represent the unified system (true values), is divided into several LHS, each associated with a specific reference tide gauge. LHS are not mutually consistent because they are not aligned to a common reference, but instead differ by systematic biases, such as offsets and tilts, collectively referred to as \texttt{"vertical datum parameters"}. Unification is then carried out with respect to a pre-selected LHS (hereafter referred to as the \texttt{"unifying datum"}), which is known to have a specific offset with respect to the IHRS reference potential $W_0$. The local heights are simulated by perturbing the true values by adding these assumed systematic biases. In this study, we consider the following vertical datum parameters: vertical datum offsets ($\delta^L$), local vertical datum alignment discrepancies in the latitude and longitude directions ($a^L$ and $b^L$), accumulated tilts depending on the distance from the reference tide gauge ($t^L$), and levelling point elevation-dependent offsets ($m^L$), which can be larger in mountainous regions where levelling measurements are sparser and more prone to errors. End-to-end simulations are carried out using the simulated local heights together with clock measurements as observations to estimate the unified heights along with the vertical datum parameters between the LHS. We propose a certain number of clocks $n$ distributed in each LHS with several connections (i.e. links) between them. The number of clock observations must exceed the total number of parameters to be determined, thereby ensuring an over-determined solution. If all clocks are interconnected, we can have a total of $\binom{n}{2}$ observations. This larger number can allow for a reduction of measurement errors related to the clock and link uncertainties, depending on the correlations between the different effects contained in the measurements. The unification accuracy is assessed through the adjusted height residuals, defined as the differences between the unified heights after adjustment and their a priori heights. These residuals reflect how effectively the adjustment based on clock measurements has reduced the true height residuals in the simulated local heights.

This study extends the work of \cite{wu2018} in several novel ways:
\begin{itemize}
\item[\textbullet]Use of an extended parametric model to determine the vertical datum unification parameters between LHS to reflect more realistic and complex conditions. While \cite{wu2018} considered $\delta^L$, $a^L$ and $b^L$, we augment the model with two further parameters: $t^L$ and $m^L$;
\item[\textbullet]Optimization of the distribution of clocks within each LHS to improve unification accuracy;
\item[\textbullet]Application to a new test region -- Brazil -- in addition to Europe;
\item[\textbullet]Realistic modelling of clock observations considering:
\begin{itemize}
    \item[-]Tidal effects and their removal from clock observations, including model errors
    \item[-]Clocks with different frequency uncertainties within each LHS;
    \item[-]Realistic clock link uncertainties, accounting for both optical fiber and free-space signal propagation;
    \item[-]Analysis of the influence of outliers in clock networks with fewer connections (i.e. observations);
    \item[-]Impact of correlations between clock observations on the unification process;   
\end{itemize}
\item[\textbullet]Discussion on how the resulting unified height system can be related to the global geoid.
\end{itemize}

The research focus differs between the Europe-only and Brazil-only cases. In Europe, the study concentrates on realistic and complex unification parameters between LHS, the optimal number and spatial distribution of clocks, and unification with different frequency standards. The number of clock links and their uncertainties, outlier analysis, and temporal correlations are the focus in Brazil.

\subsection{Data Preparation}\label{sec:data}
To simulate clock observations and local height observations, we require either the direct value of $C_p$ of the levelling points in the height systems or the normal heights $H^N = \frac{C_p}{\overline{\gamma}}$ as the a priori true values. $\overline{\gamma}$ is the mean normal gravity value along the normal plumb line associated with the level ellipsoid \citep{jekeli2000}. The European Vertical Reference Network (EUVN) height solution, for which the vertical datum is related to NAP (Normaal Amsterdam Peil), provides normal heights all over Europe \citep{ihde2000}. With the computed value of $\overline{\gamma}$, $H^N$ of each levelling point is transformed to the corresponding potential number $C_p$. Brazilian levelling data in terms of $C_p$ were obtained from SIRGAS, the Geocentric Reference System for the Americas, with the Imbituba tide gauge selected as the unifying datum, as its difference to the IHRS $W_0$ value presents the least standard deviation.

\subsection{Extended Model for Vertical Datum Parameters}\label{sec:bias}
The vertical datum parameters are determined by modelling the biases using a parametric function with one, two, three or more parameters. Hence, local height values ($H_{i}^L$) in terms of geopotential numbers ($C_{i}^L = gH_{i}^L$) are generated by adding all of these biases and random errors ($\epsilon)$ to the a priori (true) values \(C_{i}^U\), the geopotential number referring to the unifying datum, which can be represented as
\begin{equation}
C_{i}^L = C_{i}^U + (a^{L} \Delta X_i + b^{L} \Delta Y_i + t^{L} \Delta S_i) \cdot g  + m^{L} C_{i}^U/E + \delta^L + \epsilon. \label{eq:errors}
\end{equation}
$\Delta X_i$ and $\Delta Y_i$ represent the latitude and longitude differences (in degrees), measured from the minimum latitude and longitude values in the LHS. $\Delta S_i$, is the distance from the tide gauge in degrees. We model $m^L$ as a height-dependent offset that only becomes relevant for levelling points above a certain threshold elevation, chosen here as $E = 500$~m. This threshold is selected as a practical reference scale: it is large enough to represent typical elevation changes in mountainous regions while avoiding sensitivity to small height differences. For points below 500~m, the offset is assumed negligible.
To convert the tilt effects $(a^{L} \Delta X_i + b^{L} \Delta Y_i + t^{L} \Delta S_i)$ to gravity potential a mean value of $g = 9.8$ m$/$s$^2$ is used. A random noise $\epsilon$, assumed to follow a Gaussian distribution [$\epsilon \sim \mathcal{N}(0,\sigma_\epsilon^2$,] with standard deviation $\sigma_\epsilon = 0.1~\text{m}^2/\text{s}^2$, is added to the model. Equation~(\ref{eq:errors}) represents an extended model, incorporating five vertical datum parameters, developed to account for more realistic and complex scenarios, while \cite{wu2018} conducted the study on a simplified model with three parameters, $a^L$, $b^L$ and $\delta^L$.
 
\subsection{Realistic Clock Observations}\label{sec:ClockUnc}

A clock observation $\Delta W_{ij}$ is the gravity potential difference between two sites as inferred from the redshift $z$ measured between two clocks. It is affected by clock biases ($CE$) from the two clocks that can be correlated, clock comparison error ($PD$) and external effects such as tidal effects ($TE$). The observation equation for the clock measurements ($\Delta W_{ij}$) is inferred from the a priori $C_{i}^U$,
\begin{equation}
\Delta W_{ij} = -(C_{i}^U - C_{j}^U) - (TE_{i} - TE_{j}) - (CE_i - CE_j) + PD_{ij}, \label{eq:DeltaW}
\end{equation}
where $i$ and $j$ represent the clock sites. \cite{wu2018} does not account for $TE$, $CE$ and $PD$.
We assume high-performance clocks with clock biases $CE$ that are modelled as a random but constant offset, which is unique to each clock and is fixed over one observation. It is treated as an independent and identically distributed Gaussian variable with a zero mean and a standard deviation corresponding to a clock uncertainty between $10^{-17}$ and  $10^{-18}$, i.e., 1 m$^2/$s$^2$ and 0.1 m$^2/$s$^2$ respectively (see Eq.~\ref{eq:z}). Similarly, a stochastic link error equal to the link uncertainty in the range between $10^{-17}$ (1 m$^2/$s$^2$) and  $10^{-19}$ (0.01 m$^2/$s$^2$) is used for $PD$.

\subsubsection{Clock Biases and Correlation}
In clock comparison experiments, different contributions influence the measurement result and its uncertainty. Each clock is affected by random frequency fluctuations arising from noise processes inherent to the clock and the measurement system.
This leads to statistical uncertainty and random offsets, which both reduce with averaging time and can be evaluated directly from repeated measurements or from the observed frequency stability of the clock during the comparison interval. 
These contributions become negligible with a sufficiently long averaging time. Depending on the type of optical clocks and their uncertainties, averaging times between several hours and several days are required to achieve this. We assume daily-averaged clock comparisons in the simulations.

Systematic uncertainty, in contrast, is associated with frequency shifts caused by well-identified physical effects. These effects are corrected using information from physical models or experimental investigations; however, the corrections are subject to uncertainties arising from incomplete knowledge and limitations in their characterization. Thus, the correction will, in general, deviate from the true, unknown systematic shift, and a clock bias $CE$ remains. While biases $CE$ are often stable within a single campaign, they may differ between campaigns due to changes in environmental conditions, clock configuration, or calibration procedures.
In any case, biases $CE$ will remain within a distribution governed by the uncertainty of the clock.
As optical clocks achieve considerably lower uncertainties than their microwave counterparts, the long-term variations of the clock frequency are also expected to be correspondingly smaller. The resulting clock biases may therefore be regarded as slowly varying (quasi‑static) stochastic quantities.

Generally, we assume that the systematic bias of each clock is constant within each observation, but uncorrelated between different observations involving a specific site. Treated as uncorrelated white noise, this reflects the assumption that such biases, while unpredictable for a clock observation, are equally likely to be positive or negative and can statistically cancel out. In Sec.~\ref{sec:tcorr}, we discuss a scenario for Brazil with correlated clock biases between clock observations involving common sites.

\subsubsection{Clock Comparison Uncertainties}
For the errors that originate from the clock comparison ($PD$), we consider two independent contributions: noise originating from the distribution of the clocks' frequencies to determine the redshift between two sites, i.e. from the link, and systematic errors occurring in this process. 
Which contributions to $PD$ are significant depends on the type of link used for the clock observation and its averaging time. If optical interferometric fiber links \citep{williams2008} are used to compare clocks, the problem can be strongly simplified due to negligible link noise and systematic effects. Typically, link noise reduces much faster with averaging time than the clock-related noise \citep{lisdat2016}, and the latter dominates when averaging longer than 100~s. Furthermore, the link's systematic error values in the order of $10^{-19}$ can be reached under realistic field conditions, as it was demonstrated in loop configurations \citep{kok19}. In consequence, we can neglect the link contributions compared to the systematic clock biases $CE_i$.

When no interferometric fiber links are available because of missing infrastructure or too long distances, clock connections need to be realized by free-space links, typically to space. Such links exist and employ different microwave technologies to link clocks \citep{rie20} and can reach $10^{-17}$ uncertainties \citep{fuj18, jia23b}. Optical links are under development \citep{cal23, rov16}, but not yet in use on or to satellites. These links typically exhibit higher instability than interferometric fiber links and require longer averaging times to reduce statistical noise contributions. A promising tool for global clock comparisons has now become available by the ACES mission \citep{Cacciapuoti2011}, which will allow for global frequency transfer with uncertainties as low as $\approx 5\times10^{-17}$.
We consider different cases for uncertainties of free-space links: presently realistic numbers of $10^{-17}$ as well as more optimistic ones for future optical links with $10^{-18}$ uncertainty. These uncertainties are dominating in $PD$. The applied values are given in the discussion of the respective scenario below (Sec. \ref{sec:prop}).

\subsubsection{Tidal Effects on Clock Observations and their Correction}
In a real-world scenario, clock observations are affected by the fluctuating Earth's gravity potential, including external factors like tide-generating potential and accurate reductions are required. 
\begin{figure*}[h!]
\centering
{\subfigure
{\includegraphics[width=0.495\textwidth]{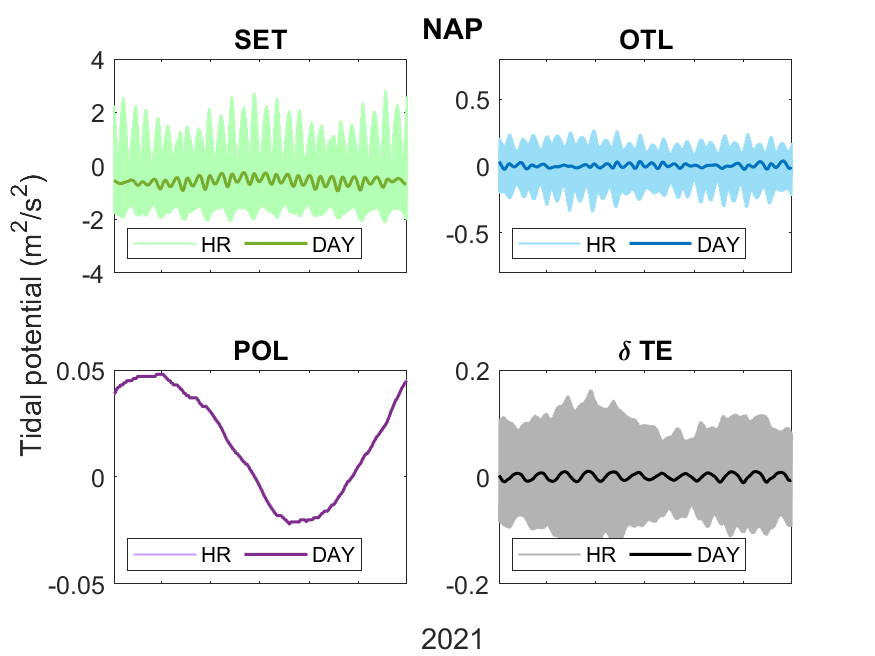}}
\subfigure
{\includegraphics[width=0.495\textwidth]{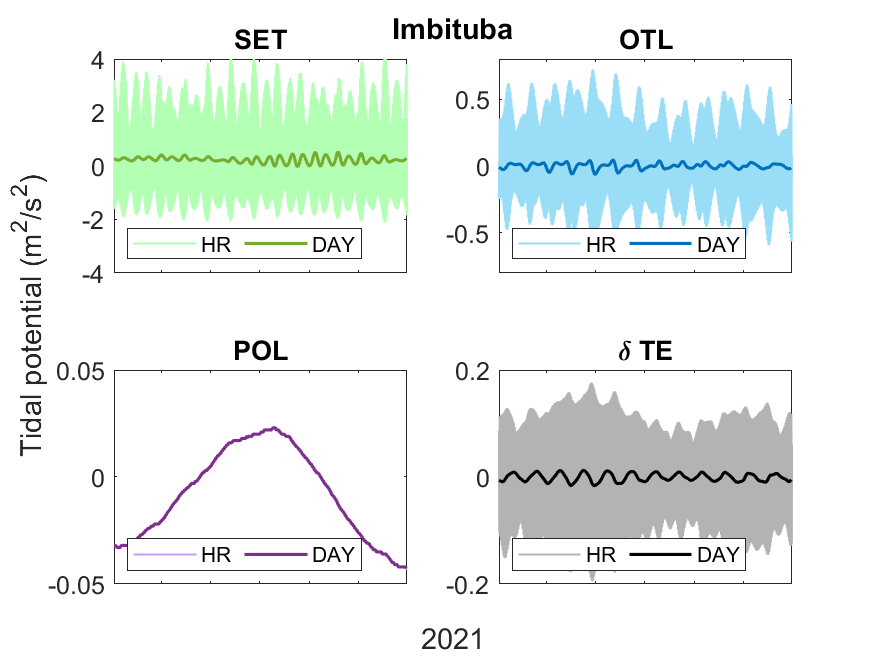}}}\\
\caption{Hourly and daily potential variations due to solid earth tide (SET), ocean load tide (OTL) and pole tide (POL), with corresponding model error ($\delta TE$) at NAP (Normaal Amsterdam Peil), the datum point of the European system (left) and Imbituba, the datum point of the Brazilian system (right).} 
\centering \label{fig:tides}
\end{figure*}
The larger tidal effects are solid-earth tide (SET) and ocean tidal loading (OTL) caused by the Sun, Moon, and other celestial objects \citep{jentzsch2005}. Centrifugal effects due to polar motion generate the pole tide (POL), while variations in Earth's angular velocity give rise to length-of-day (LOD) tides \citep{mccarthy1993}. Also, atmospheric tides \citep{lindzen1969} have to be considered. The LOD and atmospheric tidal potentials can be neglected as their range is below clock sensitivity $10^{-18}$ ($\approx$ 0.1 m$^2$/s$^2$) \citep{voigt2016}. A modified version of the ETERNA34 (PREDICT program) Earth tide data processing package \citep{wenzel2022} with the HW95 (Hartmann and Wenzel 1995 \citep{hartmann1995hw95}) tidal potential catalogue (TGP) is used to model hourly SET and POL tidal potential variations. The corresponding hourly OTL values with respect to EOT11a (Empirical Ocean Tidal Model) \citep{dgfi2012} were generated using the SPOTL3.3.0.2 package \citep{agnew2012}.

We model tidal effects $\Delta C_i(t)$ in terms of variations in $C$. An assumption was made that all clock observations were carried out in the year 2021, for which the daily tidal effects are computed by averaging the modelled hourly solutions, as clock observations are also assumed to be daily averaged. To simulate the model errors, tidal potential variations were estimated using different models of the TGP catalogue (Tamura87 \citep{tamura1991procedure}) and ocean tide model (FES2004 \citep{lyard2006modelling}). The estimated model differences ($\delta TE_i$) are added as additional errors to the tidal potential values $\Delta C_{i}(t)$
%equation
\begin{equation}
TE_i = \Delta C_{i}(t) + \delta TE_i,
\end{equation}

The tidal potential values vary depending on the latitude. Hourly tidal potential variations (SET+OTL+POL) between the clock sites are in the order of [$-4$ to 4] m$^2$/s$^2$ and average down to [$-0.4$ to 0.4] m$^2$/s$^2$ over a day. Similarly, the model error $\delta TE$ reduces from $\approx$~0.1~m$^2$/s$^2$ to $\approx$~0.02~m$^2$/s$^2$ when daily averages are computed from hourly solutions. As an example, tidal potential variations generated at the datum points of the European system (NAP) and Brazilian system (Imbituba) are shown in Fig.~\ref{fig:tides}. Short-baseline clock comparisons are less affected by tidal effects since they experience similar influences, reducing their differential impact. During the height system unification, we correct the clock observations $\Delta W_{ij}$ for the respective estimates of the tidal influences $\Delta C_{i,j}$. In consequence, the clock observations are only influenced by the model errors $\delta TE$ $\approx 10^{-19}$, hence, tidal effects can be corrected to a negligible order.

\subsection{Clock-based Adjustment}
The reunification procedure is performed through a joint least-squares adjustment of the simulated local height observations (Eq~(\ref{eq:errors})) and the simulated clock observations (Eq.~(\ref{eq:DeltaW})), which together relate all sites to a common reference datum. In our simulation, the observation matrix $\mathbf{l_{nc+nw\times1}}$ consists of the simulated local heights in terms of geopotential numbers $\mathbf{C_{nc\times1}^L}$ and clock observations simulated as gravity potential differences $\mathbf{\Delta W_{nw\times1}}$. %where $\mathbf{nc}$ and $\mathbf{nw}$ are the numbers of local height and clock observations, respectively. 
The functional model reads
\begin{equation}
\begin{bmatrix} \mathbf{C^L} \\ \mathbf{\Delta W} \end{bmatrix} = \begin{bmatrix} \mathbf{A1} & \mathbf{A2} \\ \mathbf{A3} & \mathbf{A4} \end{bmatrix} \begin{bmatrix} \mathbf{X1} \\ \mathbf{X2} \end{bmatrix}
\end{equation}
where $\mathbf{A1_{nc\times nc}=I}$ and $\mathbf{A2_{nc\times nu}}$ contain the partial derivatives of $\mathbf{C}$ with respect to $\mathbf{X1}$ and $\mathbf{X2}$ based on Eq. (\ref{eq:errors}) while $\mathbf{A3_{nw\times nc}}$ and $\mathbf{A4_{nw\times nu}=0}$ contain the partial derivatives of $\mathbf{\Delta W}$ with respect to $\mathbf{X1}$ and $\mathbf{X2}$ based on Eq. (\ref{eq:DeltaW}) are the components of the design matrix $\mathbf{A}$. The vector of unknown parameters $\mathbf{X_{nc+nu\times1}}$ is represented by $\mathbf{X1} = \{..., C_{i}^U, ...\}^{'}_{nc\times1}$, the set of geopotential numbers in the unified system, and $\mathbf{X2} = \{ a^L, b^L, \delta^L, t^L, m^L\}^{'}_{nu\times1}$, the set of vertical datum parameters from every LHS ($L$). Here, the weight matrix is $\mathbf{P} = \begin{bmatrix} \mathbf{P1} & \mathbf{0} \\ \mathbf{0} & \mathbf{P2} \end{bmatrix}$ where, $\mathbf{P1}$ and $\mathbf{P2}$ represent weighting matrices for $\mathbf{C^L}$ and $\mathbf{\Delta W}$ when there is no correlation between observations. 
Unit weights are assigned to the local height observations (constant variance, $\sigma_i^2 = 0.01$) by assuming an a priori variance factor of $\sigma_0^2 = 0.01$, hence $\mathbf{P1=I}$. Similarly, unit weights are assumed for clock observations with a total uncertainty level in the order of $10^{-18}$ ($\sigma_i^2 = 0.01$), $\mathbf{P2=I}$. In practice, weighted least squares is used to account for differing variances or correlations in the clock observations, as discussed in Sections \ref{sec:diffUnc}, \ref{sec:prop}, \ref{sec:paramest}, and \ref{sec:tcorr}. The true height residuals before unification are defined as $(C_i^L - C_i^{U,\mathrm{apriori}})/g$, and the adjusted height residuals after unification as $(\hat{C}^U_i - C^{U,\mathrm{apriori}}_i)/g$, where $C_i^{U,\mathrm{apriori}}$ denotes the true geopotential numbers in the a priori system and $\hat{C}_i^U$ denotes the estimated geopotential numbers in the reunified system.

\section{Results and Discussions}\label{sec5}
\subsection{European Height System -- Relation to NAP}\label{sec:NAP}
\begin{figure*}[h!]
\centering
\includegraphics[width=0.550\textwidth]{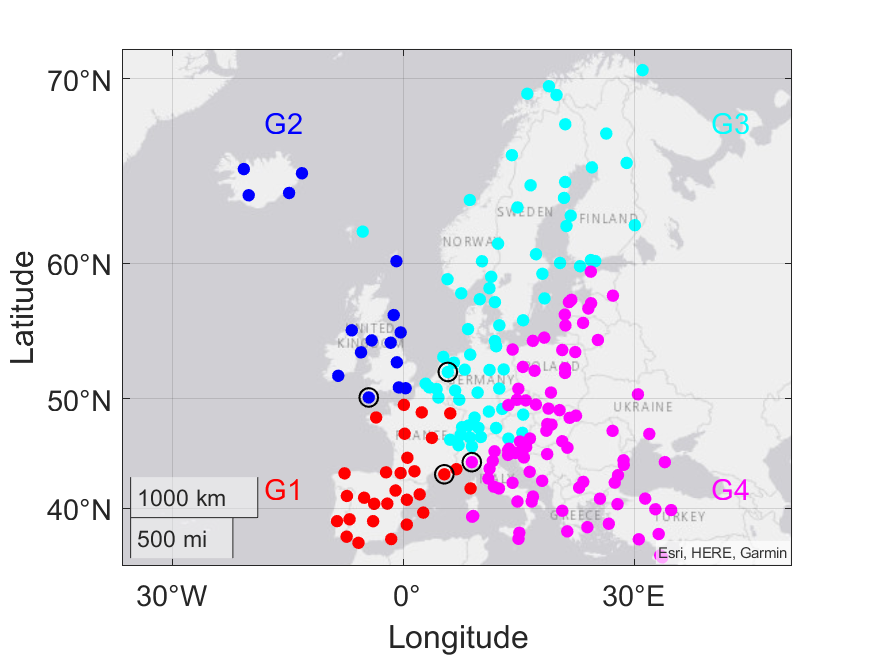}
\caption{Local height systems across Europe used in this simulation study. The European Vertical Reference Network (EUVN) is divided into four local height systems (LHS): G1, G2, G3, and G4, each represented by a different colour. The reference tide gauges for each LHS are marked by black circles.} \label{fig:LHS_NAP}
\end{figure*}
\begin{figure*}[h!]
\centering
{\subfigure
{\includegraphics[width=0.47\textwidth]{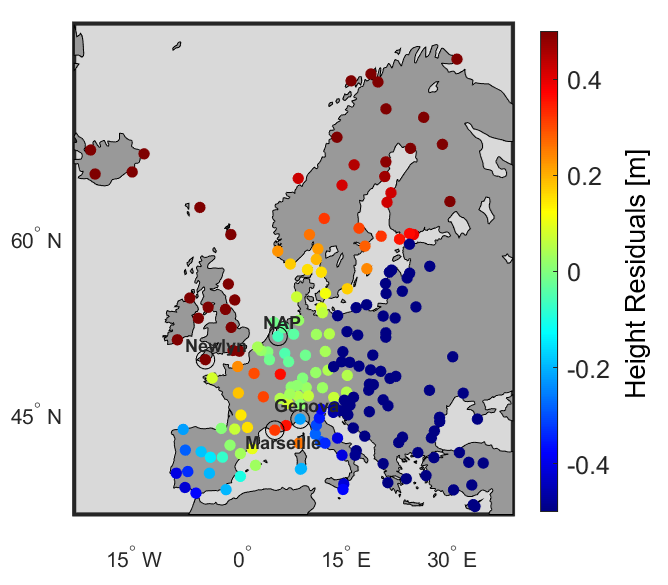}}
\subfigure
{\includegraphics[width=0.47\textwidth]{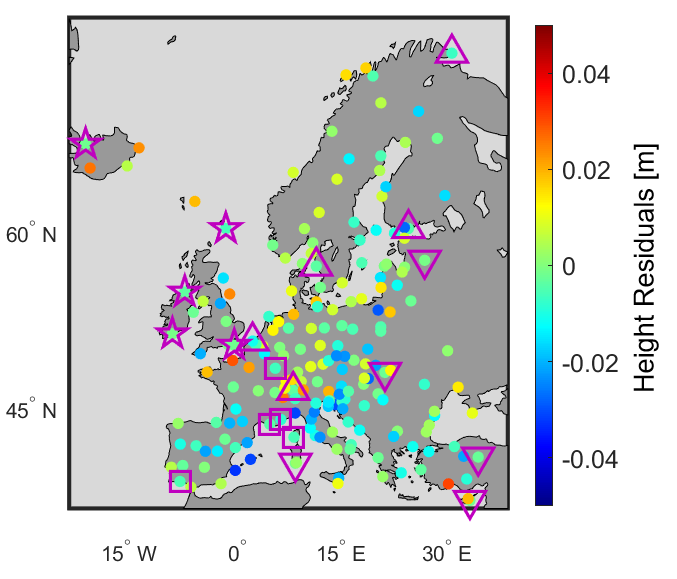}}}\\
\caption{True height residuals before unification (left) and the adjusted height residuals after unification (right) obtained with the extended model, Eq.(\ref{eq:errors}). Black labels on the left represent the simulated reference tide gauges. Different symbols on the right represent the clock sites in each LHS. The adjusted height residuals reduced  to $\approx \pm$ 1 cm after the clock-based unification.}  %%
\centering \label{fig:res_NAP}
\end{figure*}

For this simulation study, 202 levelling points from the EUVN2000 network are taken and classified into four local systems, G1, G2, G3 and G4 (Fig.~\ref{fig:LHS_NAP}). Local heights are simulated by adding the parameters as described in the extended model (Eq.~(\ref{eq:errors})). The rationale behind the added biases is based on \cite{gruber2014height} and \cite{wu2018}. We assume 5 vertical datum parameters ($a^L, b^L, \delta^L, t^L, m^L$) in each LHS, where the offset $\delta^{L3} = 0$ as the pre-defined unifying datum is NAP, the local vertical datum of G3 and $m^{L2} = 0$ because there are no high elevated points ($> 500$~m) in G2. The assumed vertical datum parameters before unification in each LHS are provided in Tab.~\ref{tab_Europe}. With five clocks of $10^{-18}$ uncertainty in each LHS and all clocks being interconnected, we have a total of 190 clock observations. The clock uncertainties between these observations are treated as uncorrelated. Unification is carried out using a least-squares adjustment with the observation equations (\ref{eq:errors}) and (\ref{eq:DeltaW}). $PD$ is in the order of $10^{-19}$ since optical fiber links are considered. The specific distribution of the clock network is discussed in detail in Sec.~\ref{sec:ClockNet}, for which the unification results are shown in Fig.~\ref{fig:res_NAP}. The difference between unified heights and a prior heights is provided as adjusted height residuals in Fig.~\ref{fig:res_NAP} (right). The overall unification accuracy for each LHS is in the order of 1~cm (Tab. \ref{tab:rms1}).
\begin{table*}[h!]
    \centering
    \caption{Vertical datum parameters assumed in the 4 LHS of the European system before unification}
    \begin{tabular}{|p{2.5cm}|p{0.6cm}|p{0.6cm}|p{0.6cm}|p{0.6cm}|}
    \hline
         \centering
         Vertical datum parameters& \textcolor{red}{G1}& \textcolor{blue}{G2}& \textcolor{cyan}{G3}& \textcolor{magenta}{G4}\\
         \hline
         $m^L$ (cm)& 1.5& 0& 2& 1.5\\
         \hline
         $t^L$ (cm/degree)& $-2$& 3.5& 3& $-3$\\
         \hline
         $a^L$ (cm/degree)& 3& $-2$& 1.5& $-3$\\
         \hline
         $b^L$ (cm/degree)& 2& 3& $-1.5$& $-2$\\
         \hline
         $\delta^L/g$ (cm)& $-18$& 25& 0& 8\\
         \hline
    \end{tabular}
    \label{tab_Europe}
\end{table*}
\begin{table*}[h!]
    \centering
    \caption{RMS of the adjusted height residuals of each LHS -- \textcolor{red}{G1}, \textcolor{blue}{G2}, \textcolor{cyan}{G3} and \textcolor{magenta}{G4} after unification with $10^{-18}$ clocks in Europe}
    \begin{tabular}{|p{3cm}|p{0.6cm}|p{0.6cm}|p{0.6cm}|p{0.6cm}|}
    \hline
         \centering
         LHS & \textcolor{red}{G1}& \textcolor{blue}{G2}& \textcolor{cyan}{G3}& \textcolor{magenta}{G4}\\
         \hline
         RMS (cm)& 1.46& 1.37& 1.09& 1.33\\
         \hline
    \end{tabular} \label{tab:rms1}
\end{table*}

\subsubsection{Clock Network Configuration}\label{sec:ClockNet}
For the estimation of the vertical datum unification parameters of each LHS, the spatial distribution of clocks plays a crucial role. Specifically, to accurately estimate tilt values, clocks should be strategically placed at both the maximum and minimum points of the tilt within the local height system.
\begin{figure*}[h!]
\centering
\includegraphics[width=0.7\textwidth]{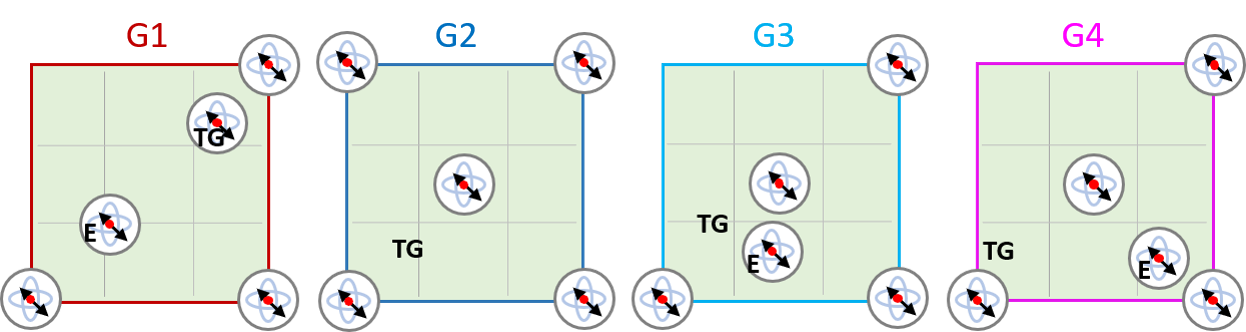}
\caption{Spatial distribution of 5 clocks simulated for each LHS -- \textcolor{red}{G1}, \textcolor{blue}{G2}, \textcolor{cyan}{G3} and \textcolor{magenta}{G4}. The preferred clock positions are at the four corners, tide gauge (TG), maximum elevated point (E) and the centre of the area covered by each LHS.} 
\label{fig:clock_distrib}
\end{figure*}
\begin{figure*}[h!]
\centering
\includegraphics[width=0.7\textwidth]{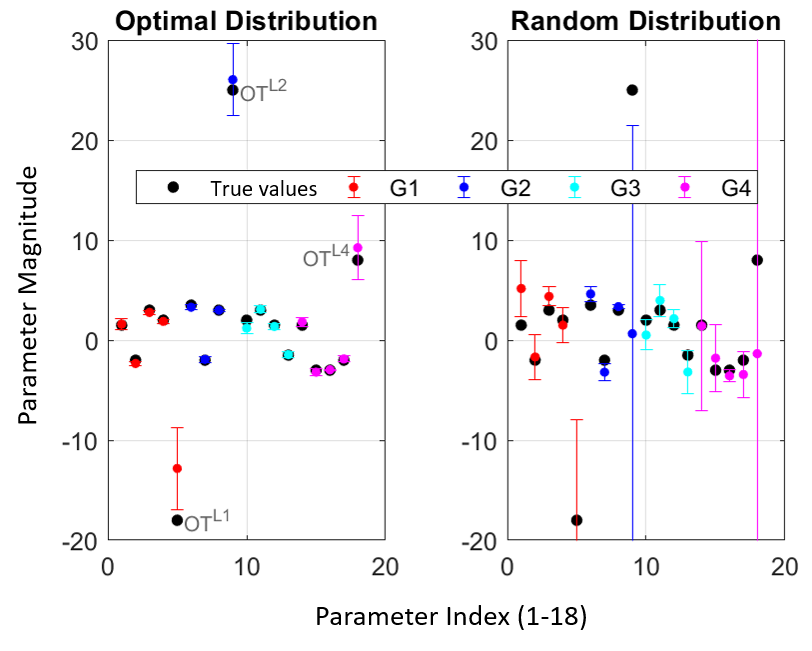}
\caption{Estimated vertical datum parameters and their standard deviations using an extended model with 18 unknowns for an optimized (left) and a random, non-optimized (right) clock distribution. 
The assumed true values of the parameters are provided for reference (Tab.~\ref{tab_Europe}). The units are different for tilts (cm/degree) and offsets (cm).}
\label{fig:stdev}
\end{figure*}
 \begin{figure*}[h!]
\centering
\includegraphics[width=0.99\textwidth]{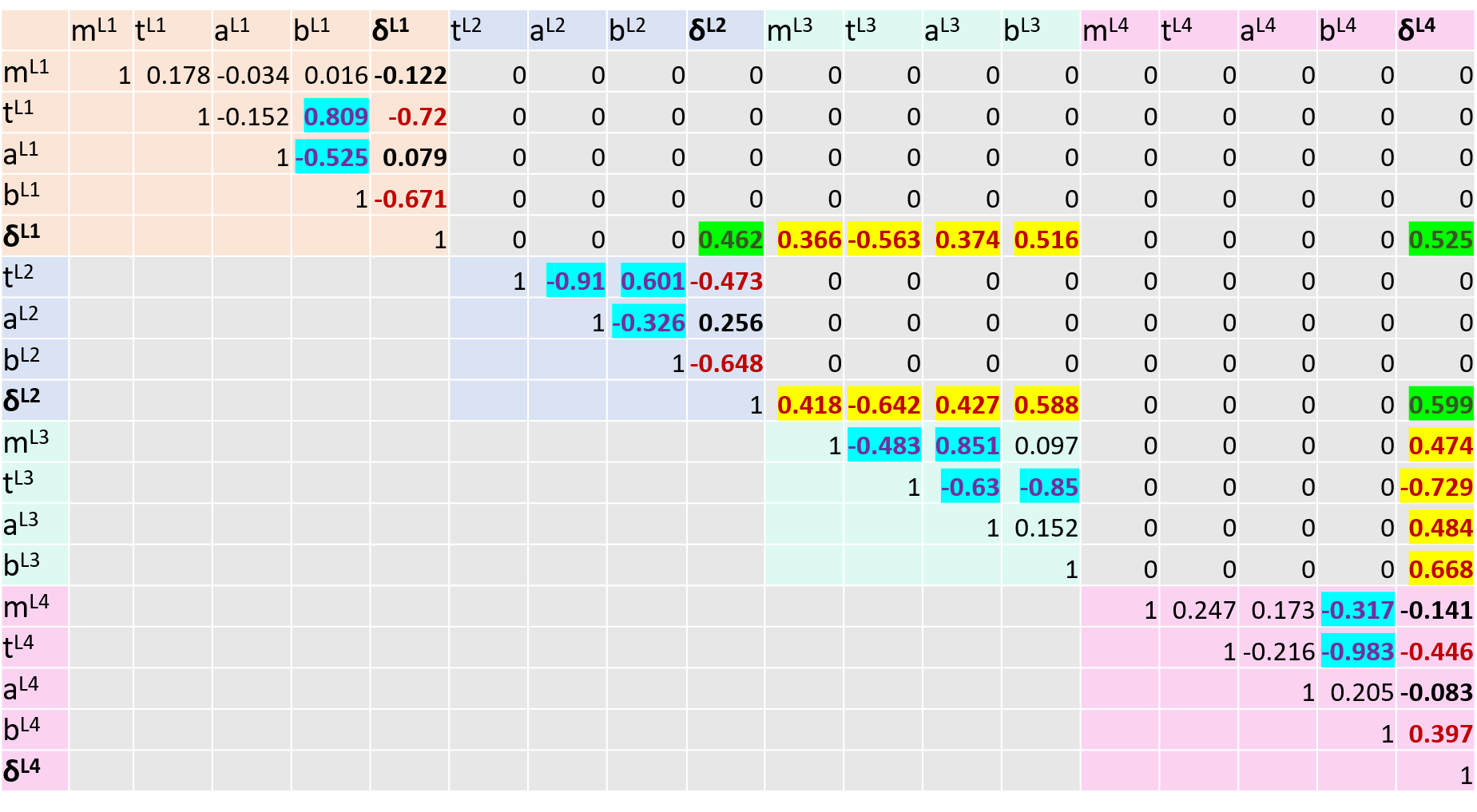}
\caption{Correlation matrix of estimated vertical datum parameters using the optimal clock distribution (see Fig. \ref{fig:clock_distrib}). The correlation between different LHS is highlighted in green and yellow. Larger correlation values within each LHS are marked in blue and red.}
\label{fig:er_corr_mat}
\end{figure*}
Conversely, for precise offset estimation, clocks should be positioned at points least affected by tilts. To accurately estimate the tilt values $a^L$ and $b^L$, clocks should be placed typically at the most northerly, southerly, easterly and westerly points of each LHS, where the direction-dependent tilt terms $a^L\Delta X$ and $b^L\Delta Y$ reach their extrema. A clock near or at the reference tide gauge site can help in estimating $t^L$ more accurately. A clock at the highest elevated point in the LHS, where $m^LC^U/E$ is maximum, allows an accurate estimation of $m^L$. Thus, the optimal spatial arrangement of clocks is contingent upon the systematic tilts and offsets present within the LHS. The idea is that tilts are direction-dependent, and clocks can be placed at the most extreme points to capture them. One can also get some a priori knowledge of these distortions through GNSS/levelling and geoid comparisons \citep{fotopoulos2003accurately}.

After analyzing the covariance matrices obtained in several simulations with random distribution of the clocks across the LHS, we select the spatial distribution shown in Fig.~\ref{fig:clock_distrib} as the optimal one. This configuration, which provides the minimal standard deviations of the estimated vertical datum parameters ($a^L, b^L, t^L, m^L, \delta^L$), is suggested for real clock measurement campaigns. Figure \ref{fig:stdev} shows the standard deviations obtained with an optimal clock distribution (left) and a random, non-optimised clock distribution (right). The largest standard deviation values obtained with the optimal configuration are $\approx$ 5 cm for the parameter $\delta^L$. For the tilts (\(a^L, b^L, t^L\)) the values are below $\approx$ 1 cm/degree and $\leq$ 1 cm for \(m^L\).
To investigate the relatively large standard deviation of $\delta^L$ and other significant correlations, a correlation matrix is computed for the optimal clock distribution. The only parameter that is correlated with other vertical datum parameters between different LHS is $\delta^L$. It shows a moderate correlation of ($\approx0.5$) with other offsets and a similar correlation with all vertical datum parameters of G3 (i.e., the unifying datum, $\delta^{L3} = 0$). Hence, the relatively large standard deviations of $\delta^L$ can be attributed to its correlations with other parameters. Within an individual LHS, $\delta^L$ is moderately correlated (ranging from 0.4 to 0.7) with $b^L$ and $t^L$. The other larger correlations ($> 0.7$) are between the tilts $b^L$ and $t^L$ within the LHS, caused by the position of the tide gauge (TG; see Fig.~\ref{fig:clock_distrib}) in the LHS. For example, in G4, TG is closer to the southwest side, causing both tilt effects to vary similarly. Thus, the strong correlation (close to 1) indicates poor parameter identifiability due to the LHS geometry. Nevertheless, the unified heights, which are of primary interest, remain accurately estimated, as shown in Section 4.1.

\subsubsection{Clocks with Different Frequency Uncertainties}\label{sec:diffUnc}
The simulations presented in Sec. \ref{sec:ClockNet} are based on only high-performance clocks with biases of $10^{-18}$.
\begin{figure*}[h!]
\centering
{\subfigure
{\includegraphics[width=0.47\textwidth]{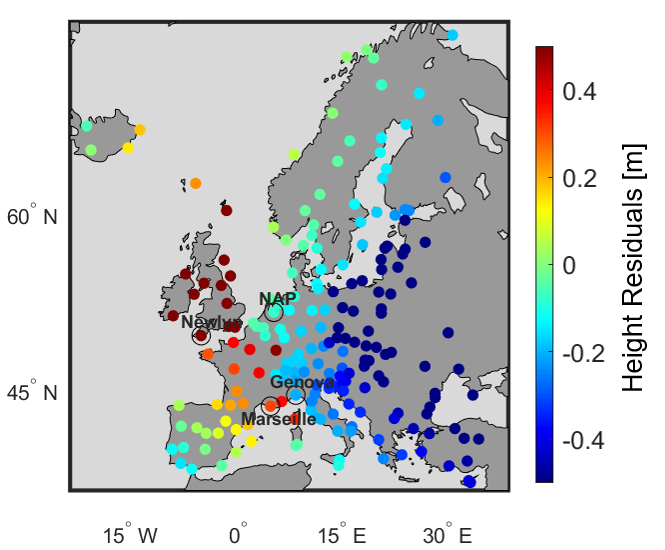}}
\subfigure
{\includegraphics[width=0.47\textwidth]{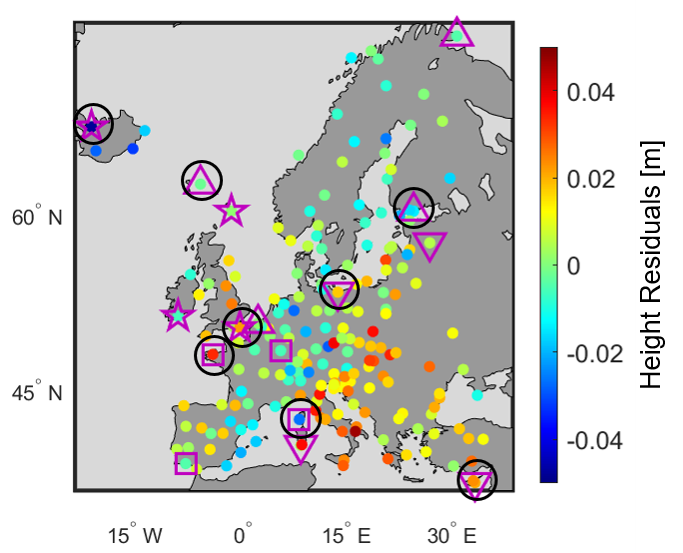}}}\\
\caption{True height residuals before unification (left) and adjusted height residuals after unification (right) when clocks with different frequency biases ($10^{-17}$ and $10^{-18}$) are used in each LHS. The more accurate clocks (10$^{-18}$) are placed at the north-east and south-west extremities of the LHS. Different symbols in the right image represent the clock sites in each LHS, where the less accurate clock sites are marked in black circles. The adjusted height residuals reduced to $\approx \pm$ 1-2 cm after clock-based unification.} %%
\label{fig:res_NAP2}
\centering
\end{figure*}
In this section, we consider an ensemble of clocks with different biases across each LHS, namely, two $10^{-17}$ clocks and two $10^{-18}$ clocks. The vertical datum parameter model (Eq.~(\ref{eq:errors})) is applied, considering a simplified model including only three biases: $a^L$, $b^L$, and $\delta^L$ (other parameters are set to zero). Four clocks are distributed at the extremities of the LHS based on the optimal distribution as described in Sec.~\ref{sec:ClockNet}. Clock observations are generated using Eq.~(\ref{eq:DeltaW}), with $PD$ in the order of $10^{-19}$ as all clocks are linked by fiber. A weighted least-squares approach is adopted, where observations with an uncertainty of $10^{-17}$ are down-weighted from 1 to 0.01 as the variance of $10^{-17}$ observations is 1 ($\sigma_i = 1$~m$^2/$s$^2$), while for unit-weight observations ($10^{-18}$) it is 0.01 ($\sigma_0 = 0.1$~m$^2/$s$^2$, standard deviation of unit-weight observations). The unification accuracy is better when the more accurate clocks ($10^{-18}$) are placed at the north-east and south-west extremities of the LHS. The comparison of the overall accuracy associated with the diagonal (I) and off-diagonal (II) distribution of more accurate clocks is provided in Tab. \ref{tab:rms_fr}. This difference can be associated with the definition of the assumed vertical datum parameters, e.g., $a^L$ and $b^L$ are defined to increase from the minimum latitude and longitude of the LHS. Hence, in this case, the north-east and south-west extremities of the LHS possess the minimal and maximal tilt effects in the simplified model. The clock at the south-west extreme point of the LHS is mostly affected by an offset, possibly indicating better efficiency in estimating the vertical datum parameters. 
The parameters $a^L$ and $b^L$ are well estimated for both distributions, whereas $\delta^L$ is determined with an accuracy of $\leq$ 2 cm for the diagonal distribution and $\leq$ 5 cm for the off-diagonal distribution.
The true height residuals before unification and the adjusted height residuals after unification are shown in Fig.~\ref{fig:res_NAP2}. Note that a unification accuracy of 1-2 cm can still be achieved with a network combination of $10^{-18}$ and $10^{-17}$ clocks in each LHS if a sufficient number of uncorrelated clock observations are available.
\begin{table*}[h]
    \centering
    \caption{RMS of the adjusted height residuals of each LHS -- \textcolor{red}{G1}, \textcolor{blue}{G2}, \textcolor{cyan}{G3} and \textcolor{magenta}{G4} after unification when clocks with different frequency uncertainties ($10^{-17}$ and $10^{-18}$) are used in each LHS. Results are shown when more accurate clocks are placed diagonally (I) and off-diagonally (II).}
    \begin{tabular}{|p{3cm}|p{0.6cm}|p{0.6cm}|p{0.6cm}|p{0.6cm}|}
    \hline
         \centering
         LHS & \textcolor{red}{G1}& \textcolor{blue}{G2}& \textcolor{cyan}{G3}& \textcolor{magenta}{G4}\\
         \hline
         RMS I (cm)& 1.38& 2.10& 1.09& 1.95\\
         \hline
         RMS II (cm)& 2.13& 3.05& 4.07& 3.13\\
         \hline
    \end{tabular} \label{tab:rms_fr}
\end{table*}

\subsection{Brazilian height system -- Unify to Imbituba}\label{sec:braz}
We have access to a total of 71\,196 geopotential numbers of Brazilian levelling data.
\begin{figure*}[h!]
\centering
\includegraphics[width=0.55\textwidth]{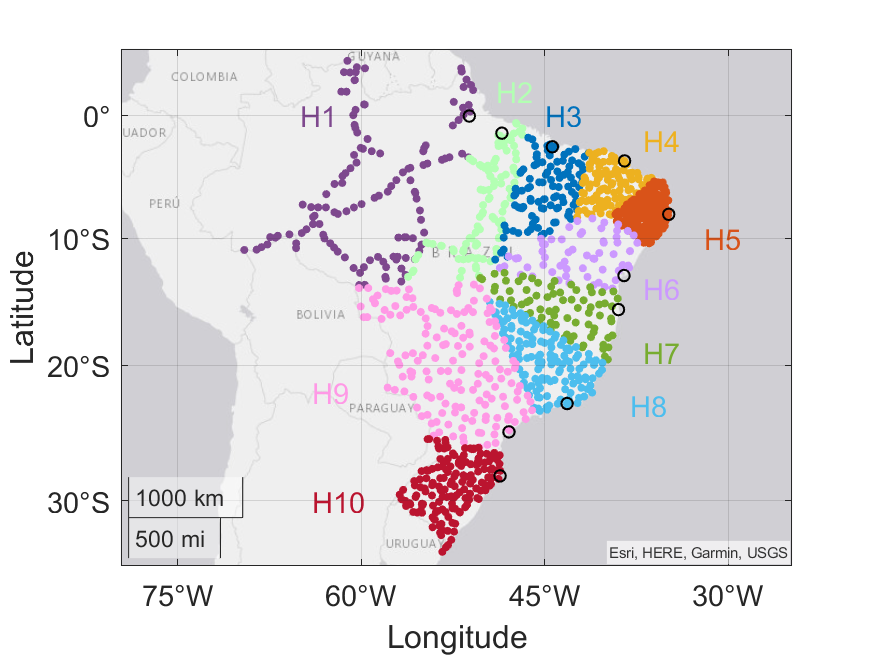}
\caption{Local height systems across Brazil used in this simulation study. The 10 LHS: H1, H2, H3, H4, H5, H6, H7, H8, H9, and H10 are indicated in different colours. The reference tide gauge locations for each LHS are indicated by a black circle.}
\label{fig:LHS_Braz}
\end{figure*}
To simplify the simulation, the number of levelling points is reduced by applying different grid sizes for each LHS. This selection process yields 1\,164 geopotential numbers. The reduced dataset is sufficient to represent Brazil’s topography accurately. Based on the tide gauge coordinates obtained from PSMSL (Permanent Service for Mean Sea Level) \citep{holgate2013new}, all points are classified into 10 LHS by assigning a reference tide gauge to each LHS, see Fig.~\ref{fig:LHS_Braz}. Thus, we have a set of 10 LHS: H1, H2, H3, H4, H5, H6, H7, H8, H9, and H10 which have to be unified with the datum of H10, Imbituba as reference.
\begin{table*}[h!]
    \centering
    \caption{Vertical datum parameters assumed in the 10 LHS of the Brazilian system before unification}
    \begin{tabular}{|p{2.5cm}|p{0.6cm}|p{0.6cm}|p{0.6cm}|p{0.6cm}|p{0.6cm}|p{0.6cm}|p{0.6cm}|p{0.6cm}|p{0.6cm}|p{0.6cm}|}
    \hline
         \centering
         Vertical datum parameters& H1& H2& H3& H4& H5& H6& H7& H8& H9& H10\\
         \hline
         $a^L$ (cm/degree)& 2& 1.5& $-2$& 1.5& 3& 2.5& 2& 2& 3.5& 3\\
         \hline
         $b^L$ (cm/degree)& 3& 2& 2& $-1.5$& 2.5& $-3$& 1.5& 2.5& 2& 3\\
         \hline
         $\delta^L/g$ (cm)& $-34$& 6& 22& $-28$& 30& 26& 15& $-10$& 35& 0\\
         \hline
    \end{tabular}
    \label{tab_Braz}
\end{table*}
Here, the height observation equation (Eq. (\ref{eq:errors})) is applied in a simplified version including only three parameters ($a^L$, $b^L$, and $\delta^L$) in each LHS, resulting in a total of 29 unknown vertical datum parameters. In this case, $\delta^{L10} = 0$. The assumed vertical datum parameters in each LHS are provided in Tab.~\ref{tab_Braz}. 
A network of 4 clocks with uncertainties of $10^{-18}$ is distributed at the four corners—north, south, east, and west—of the area defined by the LHS, in agreement with the optimal distribution as described in Sec. \ref{sec:ClockNet}. A reduced number of connections is considered, which will be detailed in the next section.

\subsubsection{Modified Clock Links}\label{sec:modlinks}
Unlike what was discussed in \citep{wu2018} and Sec.~\ref{sec:NAP}, where all clocks are inter-connected, 
\begin{figure*}[h!]
\centering
\includegraphics[width=0.45\textwidth]{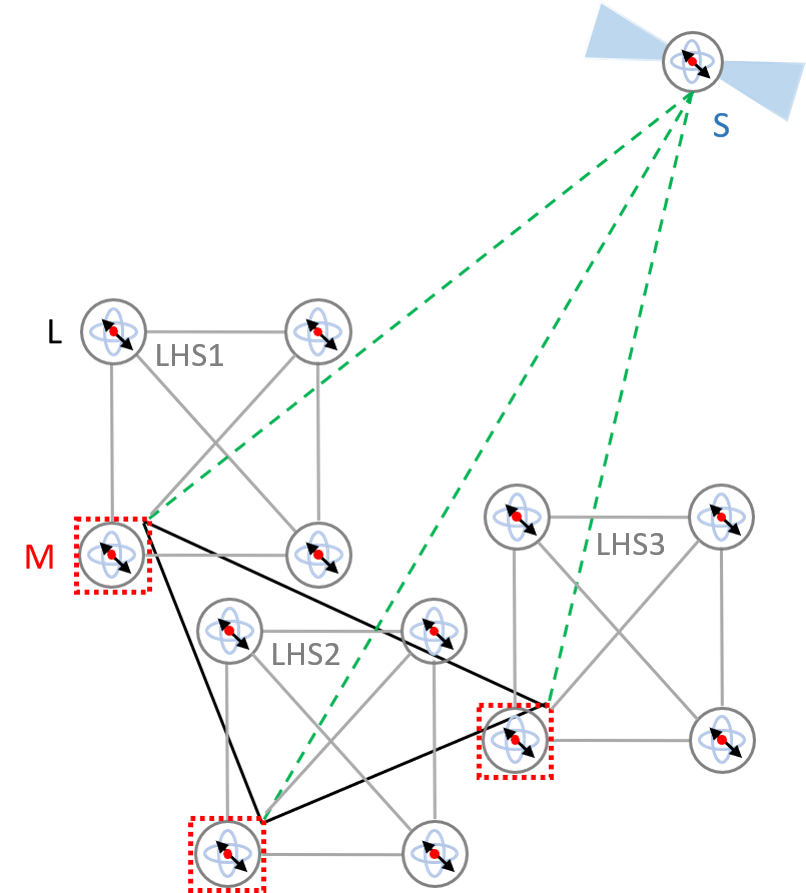}
\caption{Configuration of clock distribution with master clocks (M) and local clocks (L) in each LHS. Local clock connections are established with fiber links (grey). Master clocks can be connected via either optical fiber (black) or free-space links (green).}
\label{fig:red_links}
\end{figure*}
in this simulation case, we reduce the number of links and thus clock observations without considerably affecting the accuracy of the unification, as demonstrated in the following sections. This indicates that multiple links between the LHS may introduce redundancy without providing substantial improvement to the adjustment results. 
We assume four clocks in each LHS connected by fiber links. One of these local clocks is singled out as the master clock (i.e., all the other local clocks are connected to it), which is additionally linked to the master clocks of the other LHS by free-space links (Fig.~\ref{fig:red_links}). With four clocks in each LHS, we have 60 local links ($10 \cdot \binom{4}{2}$) and 45 master links ($\binom{10}{2}$). The connections are reduced so that there is a single connection from each LHS to any other LHS via the master clocks only.

\subsubsection{Propagation Delay in Terms of Link Uncertainty}\label{sec:prop}
\begin{figure}[h!]
\centering
{\subfigure(a)
{\includegraphics[width=0.47\textwidth]{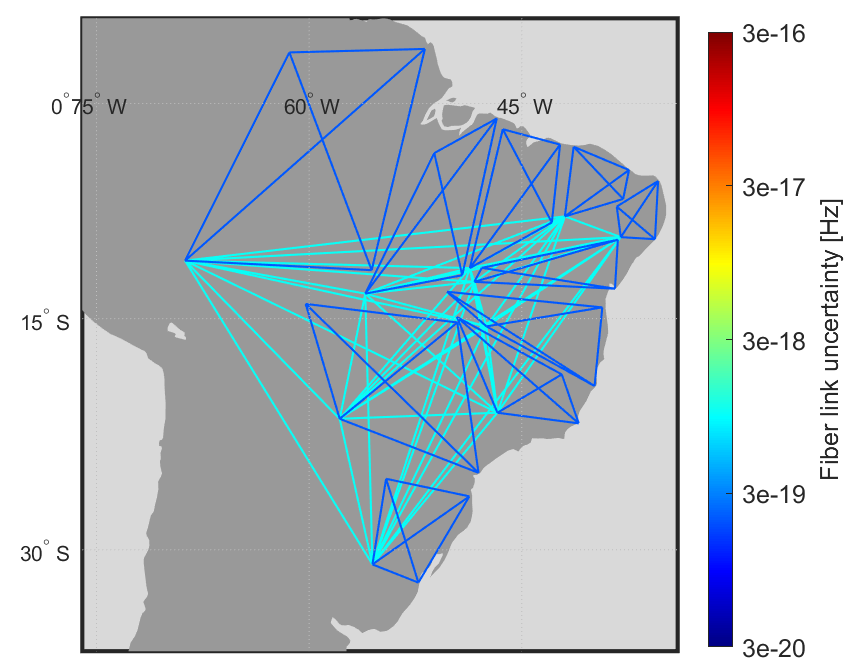}}
\subfigure
{\includegraphics[width=0.47\textwidth]{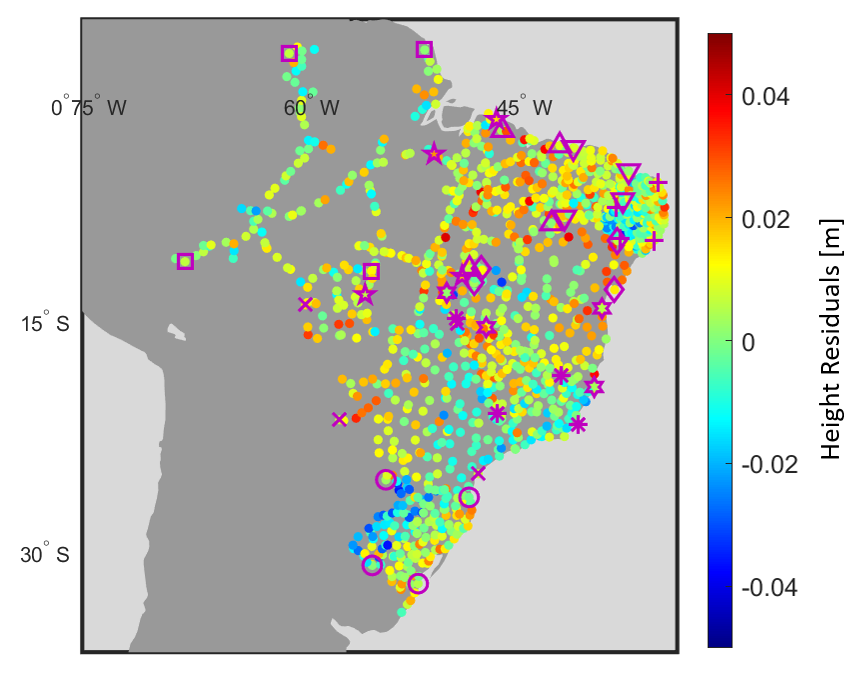}}}\\
{\subfigure(b)
{\includegraphics[width=0.47\textwidth]{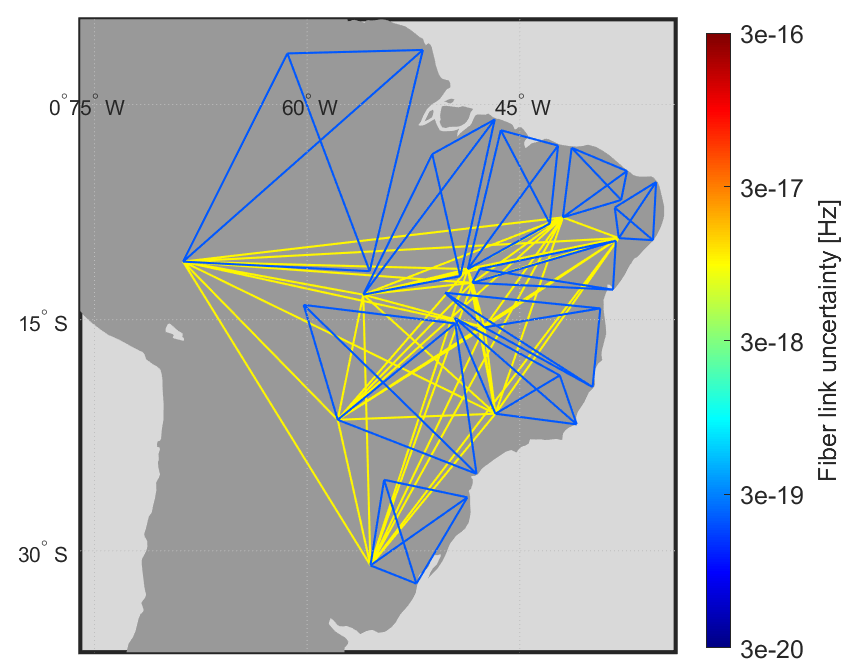}}
\subfigure
{\includegraphics[width=0.47\textwidth]{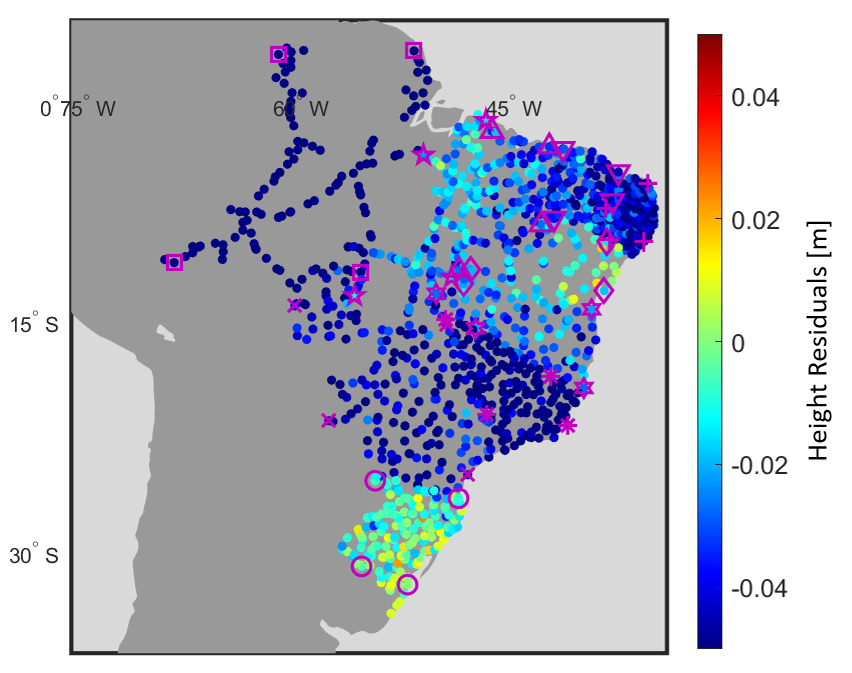}}}\\
\caption{Link uncertainty (left) and adjusted height residuals after unification (right) with different cases of free-space master clock connections. In all cases, local connections are assumed to be precise optical fiber links ($10^{-19}$). All master links are assumed as $10^{-18}$ free-space links (a) and $10^{-17}$ free-space links (b). Residuals are smaller, in the range of $\pm$ 1-2 cm, when $10^{-18}$ space links are assumed.} %%
\centering
\label{fig:res_Braz}
\end{figure}
\begin{table*}[h]
    \centering
    \caption{RMS of the adjusted height residuals of each LHS -- from H1 to H10 after unification associated with the different uncertainties on the free-space master links ($10^{-17}$ and $10^{-18}$) in Brazil}
    \begin{tabular}{|p{2.5cm}|p{0.6cm}|p{0.6cm}|p{0.6cm}|p{0.6cm}|p{0.6cm}|p{0.6cm}|p{0.6cm}|p{0.6cm}|p{0.6cm}|p{0.6cm}|}
    \hline
         \centering
         LHS & H1& H2& H3& H4& H5& H6& H7& H8& H9& H10\\
         \hline
         RMS (cm) $10^{-18}$ space link & 1.07& 1.55& 1.58& 1.3& 1.28& 1.67& 1.39& 1.35& 1.31& 1.47\\
         \hline
         RMS (cm) $10^{-17}$ space link & 6.86& 2.29& 3.27& 3.95& 4.42& 1.90& 3.03& 5.96& 4.55& 1.17\\
         \hline
    \end{tabular} \label{tab:rms_bra}
\end{table*}
In Sec.~\ref{sec:ClockUnc}, details on the formulation of expected link uncertainty during clock comparisons can be found. In our simulations, local clock connections are always assumed as optical fiber links, with uncertainty in the order of $10^{-19}$, i.e., local observations are dominated by the clock bias. Unification results with different uncertainties on the free-space master links ($10^{-17}$ and $10^{-18}$) are shown in Fig.~\ref{fig:res_Braz}. The parameter values of the LHS used in the simplified model are based on Tab. \ref{tab_Braz}. $10^{-17}$ master link observations are down-weighted from 1 to 0.01 as the standard deviation of the unit-weight master links ($10^{-18}$) is 0.1 m$^2/$s$^2$. As also shown in Fig.~\ref{fig:res_Braz}, a unification accuracy of $\approx \pm$ 1-2 cm can be achieved when $10^{-18}$ space links and $10^{-19}$ fiber links are employed in the specified link configuration. The overall accuracy in terms of RMS of the adjusted height residuals in each LHS is provided in Tab. \ref{tab:rms_bra} for the two cases. Having only $10^{-17}$ free-space link uncertainty is not sufficient for better unification, as the adjusted height residuals of most of the LHS lie in the range of $\approx$ 5 cm.
While $a^L$ and $b^L$ are robustly estimated in both cases, the accuracy of $\delta^L$ improves from $\leq$ 4 cm for $10^{-17}$ links to $\leq$ 2 cm for $10^{-18}$ links.
Since H10 is fixed as the unifying datum ($\delta^{10}=0$), its offset is not estimated. Hence, the RMS of the reduced number of adjusted height residuals of  H10 for $10^{-17}$ space links (Tab. \ref{tab:rms_bra}) is considerably smaller than the RMS of the adjusted height residuals of the other LHS. 
The residual sign pattern in Fig.~\ref{fig:res_Braz}(b) is not systematic and arises from the stochastic nature of the simulation, particularly the $10^{-17}$ master link noise. It reflects a single random realization, and different runs produce different positive–negative distributions. However, the overall statistical characteristics of the residuals remain stable across simulations.

\subsubsection{Robust Parameter Estimation Approach}\label{sec:paramest}
\begin{figure*}[htb]
\centering
%{\subfigure(a)
%{\includegraphics[width=0.45\textwidth]{adjusted_error_with_out.png}}}
%{\subfigure(b)
%{\includegraphics[width=0.45\textwidth]{adjusted_error_0wt.png}}}\\
{\subfigure
{\includegraphics[width=0.47\textwidth]{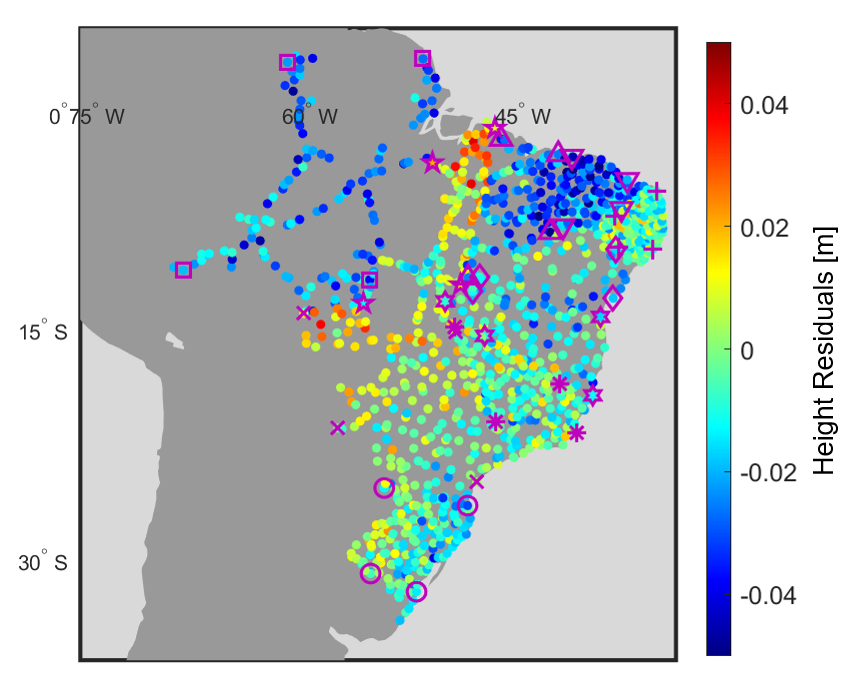}}
\subfigure
{\includegraphics[width=0.47\textwidth]{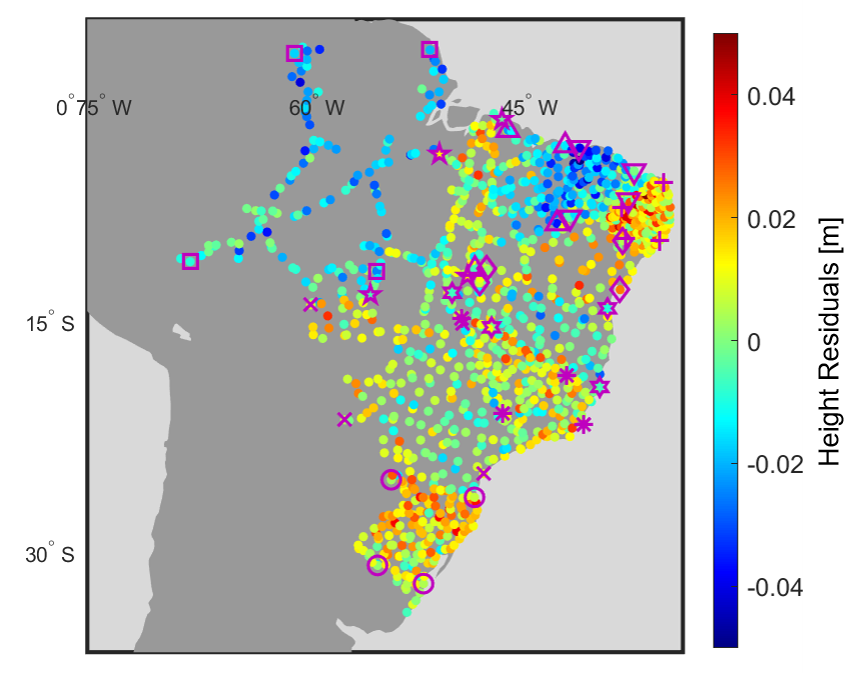}}}\\
\caption{Adjusted height residuals after unification  when Huber's robust parameter estimation \citep{Huber2011} method is applied with outliers distributed among all links (left) and only to master links (right); a unification accuracy of 1-2 cm can still be achieved if local links are outlier-free.} %%
\label{fig:out}
\centering
\end{figure*}

The possibility of outliers in real clock observations cannot be neglected. Reducing clock links to a total of 105 observations (as described in Sec. \ref{sec:modlinks}) and subsequently removing or down-weighting outlier observations can significantly impact the accuracy of height system unification, as the number of reliable clock observations diminishes. To analyze this effect, an outlier model is introduced by randomly adding outliers in the order of 50 m$^2/$s$^2$ to 5 of the 105 clock observations ($\approx 5\%$).
Outliers can be removed (weight = 0) or down-weighted (Huber's iterative reweighting algorithm \citep{Huber2011}) during the unification by means of an outlier detection method. In this case, $PD = 10^{-18}$ for free-space master links. Refer to Tab. \ref{tab_Braz} for the values assumed for parameters prior to the unification. The adjusted height residuals after unification with two scenarios, when the outlier model is applied to all links (i) and only to master links (ii), are shown in Fig.~\ref{fig:out}. In case (i), the worst unification accuracy decreased to RMS $\approx$ 3.1 cm, whereas in case (ii), it remains at RMS $\leq$ 2 cm.
$\delta^L$ estimated with an accuracy $\approx$ 5 cm for case (i) and $\approx$ 4 cm for case (ii) while $a^L$ and $b^L$ are estimated well in both cases.
This shows that down-weighting or even losing master links does not significantly degrade the results. In contrast, outliers in local links reduce the number of reliable clock observations, degrading parameter estimation and unification accuracy. In practice, reliable outlier-free local links can be ensured through highly stable fiber links. Assuming outlier-free local links, a unification accuracy of 1–2 cm remains achievable.

\subsubsection{Correlations in Clock Observations}\label{sec:tcorr}
\begin{figure*}[htb]
\centering
\includegraphics[width=0.24\textwidth]{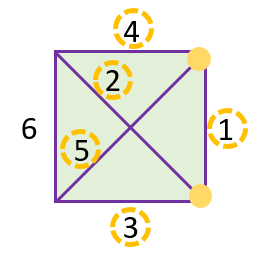}
\caption{Depiction of correlated and independent clock observations within a local height system. The observation labelled 1 is correlated with all other observations (except 6) due to common clock observations. Similarly, the other non-correlated observations are between 3 - 4 and 2 - 5.} \label{fig:clk_corr}
\end{figure*}

In this section, we analyze the impact of correlations in clock observations that are caused by clock bias contributions to $CE$ that remain (partially or fully) constant over the complete measurement campaign. Clock observations are simulated at different epochs over an assumed observation period of one year, with each observation representing a daily averaged clock comparison. Assuming there are four clocks per LHS and that each pair of clocks is connected, there are six local clock observations per LHS. However, in the case of correlated biases of a clock, correlations between observations with the same clock are strong, and the effective number of independent observations is reduced to three within the LHS (Fig. \ref{fig:clk_corr}). 
This indicates that the introduction of correlations reduces the effective information content of the observations and may degrade the unification accuracy. The adopted correlation scenarios represent simplified and controlled cases for sensitivity analysis rather than a fully physical stochastic model of clock correlations. Correlations between master links are neglected and are treated as mutually independent. $PD$ for free-space master links is assumed to be on the order of $10^{-18}$.
\begin{figure}[htb]
\centering
{\subfigure
{\includegraphics[width=0.47\textwidth]{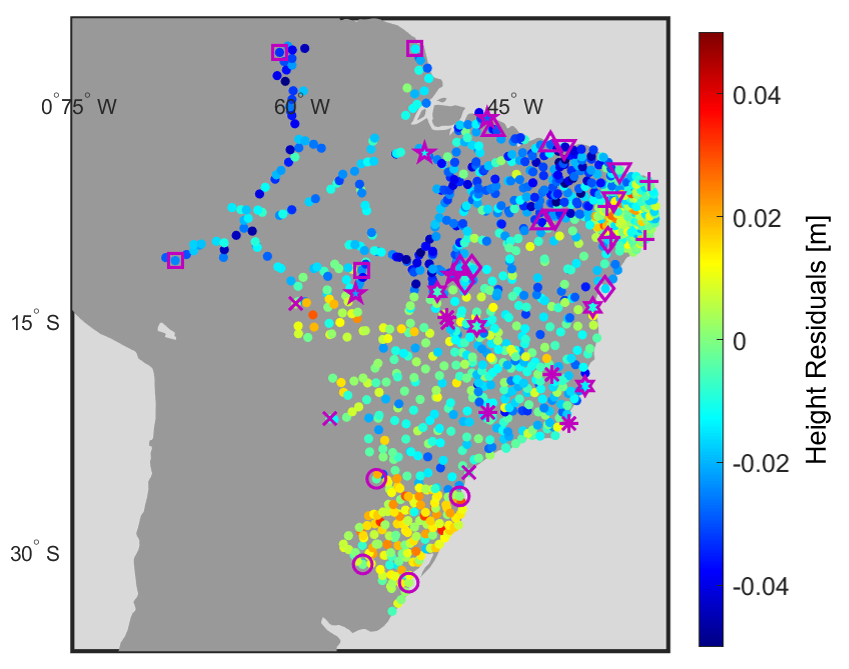}}
\subfigure
{\includegraphics[width=0.47\textwidth]{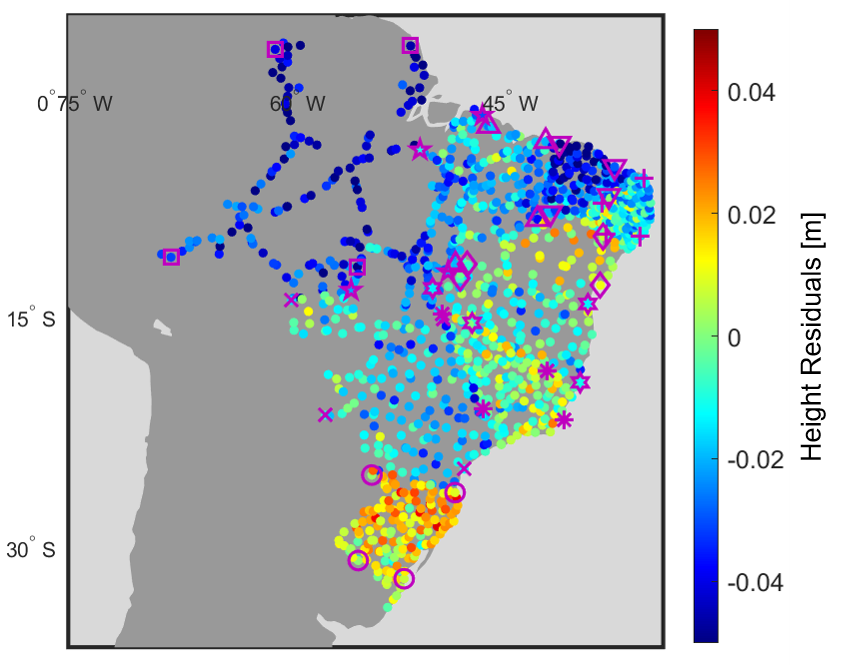}}}\\
\caption{Adjusted height residuals after unification with correlation coefficients 0.5 (left) and 0.8 (right) between the clocks; the unification accuracy remains $<$ 3 cm and $<$ 5 cm, respectively.} %%
\label{fig:tcorr}
\centering
\end{figure}

We assumed two cases: one with medium correlation ($r = 0.5$) and the second one with strong correlation ($r = 0.8$) between local clock observations. The weight matrix for the local clock observations is derived using these two correlations between observations involving the same clock. A constant variance of $\sigma^2 = 0.01$ is assumed for all clock observations. The covariance between two observations is computed as, $D_{ij} = r_{ij} \cdot \sigma_i \cdot \sigma_j$, where $-1 < r_{ij} < 1$. The weight matrix is then obtained as $P = D^{-1}$ and normalized. See Tab. \ref{tab_Braz} for the parameter values assumed in the model before unification.
There is no rank deficiency in the normal matrix. Its condition number is $2.2 \times 10^7$ without any correlation structure, increasing slightly to $3.0 \times 10^7$ for medium correlation and $4.5 \times 10^7$ for large correlation.
The adjusted height residuals after unification increase as we increase the correlation coefficient (Fig.~\ref{fig:tcorr}). The unification accuracy is $<$ 3 cm with (\(r = 0.5\)) and $<$ 5 cm with (\(r= 0.8\)). In both cases, the parameters $a^L$ and $b^L$ are well estimated. The accuracy of $\delta^L$ shows only a minor degradation, reaching $\approx$ 2 cm for $r = 0.5$ and $\approx$ 3 cm for $r = 0.8$.

\subsection{European-Brazilian Unified Height System and its Relation to the Global Geoid $W_0$}\label{sec:EU_B_geoid}
In this section, we consider the European and Brazilian height systems as two local systems, L1 and L2, with local vertical datums NAP and Imbituba, respectively. The unification of L1 to L2 involves the estimation of the offset, $\delta^{L1}$.
\begin{figure*}[h!]
\centering
\includegraphics[width=0.6\textwidth]{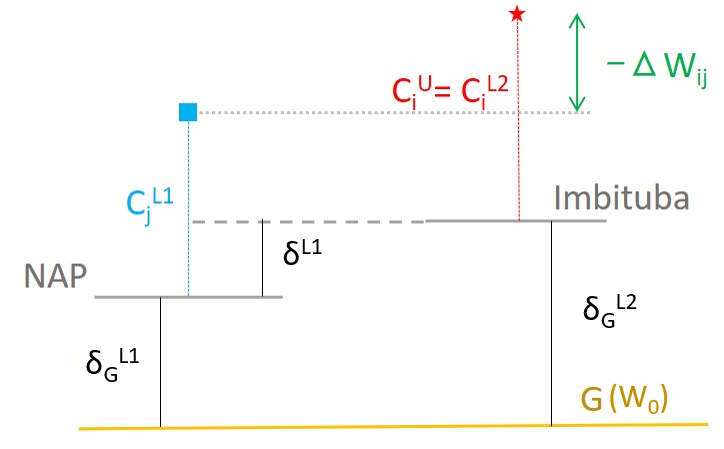}
\caption{Unification to the global geoid using a single clock observation $\Delta W_{ij}$ between site $i$ in Brazil (L2) and site $j$ in Europe (L1). The respective geopotential numbers in both systems are provided as $C_{j}^{L1}$ and $C_{i}^{L2} (= C_{i}^{U})$ where L2 is the unifying datum.} \label{fig:glo}
\end{figure*}
These vertical local datums are not aligned with the global geoid (G), i.e. the equipotential surface defined by the IHRS $W_0$ value. As the offset value of Imbituba to G ($\delta_{G}^{L2}$) is known with an estimated uncertainty of $\pm 2$ cm \citep{sanchez2017}, unifying L1 to L2 allows to refer all unified heights to G. Also, the offset between NAP to G ($\delta_{G}^{L1} = \delta^{L1} + \delta_{G}^{L2}$) can be estimated. A single clock observation between L1 and L2 is sufficient, as we have only one vertical datum parameter, $\delta^{L1}$, to estimate. However, additional observations are recommended to ensure redundancy. The relationship is illustrated in Fig.~\ref{fig:glo}.
The equation connecting the clock observation to $\delta^{L1}$ is,
\begin{equation}
    \delta^{L1} = C_{j}^{L1} - C_{i}^{L2} - \Delta W_{ij}.
    \label{eq:gg}
\end{equation}
Hence, all heights can be referenced to the geoid with a propagated uncertainty of about 3~cm, which is significantly better than the present knowledge of about 10 cm \citep{woodworth2012towards, ihde2017}.
Using Eq. (\ref{eq:gg}), this $\approx$~3 cm uncertainty results from propagating the $\approx$ 1 cm uncertainties of the two local heights and the clock observation in $\delta^{L1}$, combined with the $\pm 2$ cm uncertainty of $\delta_{G}^{L2}$.

\section{Conclusion}\label{sec13}
Our study focuses on realistic scenarios for unifying local height systems through chronometric levelling with high-performance clocks, showing the benefit of clocks in the realisation of a global height system such as the IHRS. To account for various error sources and evaluate the clock performance, we have incorporated complex vertical datum parameters into local height systems. These include datum offsets, local vertical datum alignment discrepancies in latitude and longitude directions, accumulated tilt depending on the distance from the reference tide gauge, and elevation-dependent offsets at levelling points. With the current availability of \(10^{-18}\)-level clocks, which enable a height estimation accuracy of $\approx$ 1 cm, we have investigated various uncertainties associated with clock observations and applied them in our simulation cases to assess whether the method can achieve a unification accuracy of 1 cm. Realistic daily-averaged clock observations were generated, considering the temporal correlations of intrinsic clock biases, external effects on the clock observations such as tidal effects, propagation delays in terms of link uncertainties, and outliers. Daily averaging significantly minimizes the tidal effects.

This study demonstrates that an overall $\pm$ 1-2 cm unification accuracy can be achieved by strategically placing clocks based on tilt and offset characteristics within each local height system. To achieve the best results, clocks should be placed at distant levelling points, reference tide gauges, and elevated locations, ensuring an optimal setup for the LHS unification. The study demonstrates that 1-2 cm unification accuracy can still be achieved when using a mixed network of $10^{-18}$ and $10^{-17}$ clocks. Strategic clock placement is important for improving error adjustment and achieving high-precision local height system unification. In the proposed clock network configuration with master clocks and local clocks, a unification accuracy in the range of 1-2 cm can still be achieved with fewer clock links. This reduces observational complexity without significantly compromising accuracy. Further, the outlier analysis highlights how unification accuracy is affected by whether outliers occur in master clock observations alone or across all data, underscoring the importance of accurate local clock observations linked by highly accurate optical fibers. The effect of correlations in local clock observations is analyzed, showing that adding correlation degrades the unification accuracy to a few centimetres. This underscores the importance of incorporating correlation structures in the covariance matrix using generalized least squares or other advanced statistical methods.

The European and Brazilian height systems can be unified and referenced to the global geoid with an uncertainty of $\approx$~3 cm, significantly improving the current realisations. By incorporating $10^{-18}$ clocks and similar link uncertainties, the method enables a more accurate realization of the global height reference system, enhancing the feasibility of chronometric levelling for future geodetic applications. With the development of portable clocks and proper links, short-term measurement campaigns would be sufficient for the unification process. This approach appears to be practical, reliable and more feasible in the near future.

\bmhead{Acknowledgements}
This study has been funded by the Deutsche Forschungsgemeinschaft (DFG, German Research Foundation) under Germany's Excellence Strategy EXC 2123 Quantum Frontiers -- Project-ID 90837967, SFB 1464 TerraQ -- Project-ID 434617780, and FOR 5456: Clock Metrology: A Novel Approach to TIME in Geodesy -- Project-ID 490990195.

\bmhead{Data Availability Statement}
The datasets generated and analyzed during the current study are available from the corresponding author upon reasonable request.

\bmhead{Author Contributions}
J.M. stimulated the study and provided the basic ideas. A.V. conceived the study, developed the methodology, and performed all simulations. A.V. and J.M. analyzed the results. A.V. wrote the first draft of the manuscript. C.L. provided input on the error behaviour of optical clocks. D.P. checked and optimised the relativity parts. All authors contributed to the interpretation of the data, reviewed the manuscript, and approved the final version.

\section*{Declarations}
We thank Dr.\ Eng.\ Gabriel Guimar\~{a}es, chair of the SIRGAS Working Group III (Vertical Datum), for providing the Brazilian levelling data.

\bmhead{Statement on AI Use}
The authors did not make use of AI during the preparation of this manuscript.

\bmhead{Conflict of Interest}
The authors have no conflict of interest to declare.

%\bibliographystyle{abbrvnat}
%\bibstyle{sn-basic}
\bibliography{sn-bibliography}% common bib file

\end{document}